\documentclass{svproc}
\usepackage{url}

\usepackage[dvipdfmx]{graphicx}        

\begin{document}
\mainmatter              
\title{High-Performance Cloud Computing for Exhaustive Protein--Protein Docking}
\titlerunning{Exhaustive Protein--Protein Docking on HPC Cloud}  
%
\author{Masahito Ohue\inst{1,3} \and Kento Aoyama\inst{2,4} \and Yutaka Akiyama\inst{2,3}}
\authorrunning{Masahito Ohue et al.} 
%
%
\institute{Department of Computer Science, School of Computing, Tokyo Institute of Technology, G3-56, 4259 Nagatsutacho, Midori-ku, Yokohama City, Kanagawa 226-8501, Japan\\ \email{ohue@c.titech.ac.jp}
\and
Department of Computer Science, School of Computing, Tokyo Institute of Technology, W8-76, 2-12-1 Ookayama, Meguro-ku, Tokyo 152-8550, Japan\\ \email{\{akiyama@, aoyama@bi.\}c.titech.ac.jp}
\and
Ahead Biocomputing, Co. Ltd., Kawasaki Frontier Bldg. 4F, 11-2 Ekimaehoncho, Kawasaki-ku, Kawasaki City, Kanagawa 210-0007, Japan\\ \email{\{ohue, akiyama\}@ahead-biocomputing.co.jp}
\and
 RWBC-OIL, National Institute of Advanced Industrial Science and Technology, 1-1-1 Umezono, Tsukuba, Ibaraki 305-8560, Japan}

\maketitle              

\begin{abstract}
Public cloud computing environments, such as Amazon AWS, Microsoft Azure, and the Google Cloud Platform, have achieved remarkable improvements in computational performance in recent years, and are also expected to be able to perform massively parallel computing.
As the cloud enables users to use thousands of CPU cores and GPU accelerators casually, and various software types can be used very easily by cloud images, the cloud is beginning to be used in the field of bioinformatics. 
In this study, we ported the original protein--protein interaction prediction (protein--protein docking) software, MEGADOCK, into Microsoft Azure as an example of an HPC cloud environment. 
A cloud parallel computing environment with up to 1,600 CPU cores and 960 GPUs was constructed using four CPU instance types and two GPU instance types, and the parallel computing performance was evaluated. 
Our MEGADOCK on Azure system showed a strong scaling value of 0.93 for the CPU instance when {\sf H16} instance with 100 instances were used compared to 50, and a strong scaling value of 0.89 for the GPU instance when {\sf NC24} instance with 20 were used compared to 5.
Moreover, the results of the usage fee and total computation time supported that using a GPU instance reduced the computation time of MEGADOCK and the cloud usage fee required for the computation.
The developed environment deployed on the cloud is highly portable, making it suitable for applications in which an on-demand and large-scale HPC environment is desirable.
\keywords{cloud computing, Microsoft Azure, GPU computing, protein--protein docking, MEGADOCK}
\end{abstract}
\section{Introduction}
The cloud computing environment is regarded as an important computing resource in large-scale data analysis \cite{Hashem2015,Tudoran2012}. The cloud computing environment is often used for calculation and analysis accompanied by big data, such as genomics and biomedicine \cite{Driscoll2013,Sobeslav2016}. The development of public clouds such as Microsoft Azure, Amazon AWS, and the Google Cloud Platform has contributed to the performance of large-scale bioinformatics analysis on the cloud environment \cite{Karlsson2012,Shanahan2014,Ekanavake2011}. Bioinformatics problems including sequence homology searches (BLAST and others) \cite{Matsunaga2008,Lu2010,Gunarathne2011}, similarity searches of tertiary protein structures \cite{Mrozek2014,Mrozek2016}, {\it ab initio} tertiary protein structure prediction \cite{Mrozek2015}, quantitative structure--activity relationship modeling \cite{Moghadam2015}, and protein--ligand docking \cite{Farkas2015,Paris2015} are applied in cloud computing environments as a computing resource.

Among the numerous merits of several existing cloud computing platforms, the pay-as-you-go concept whereby a user can use as much as he/she wishes at any time is the greatest advantage. Large-scale parallel computing using supercomputers enables large-scale simulation and processing of substantial amounts of data, but a user account approval procedure is required according to the institutional rules or the services are available only for the member of the organization possessing the supercomputer. In particular, several barriers exist to use for commercial purposes and owing to factors such as publicness, security and national strategy in supercomputer at public institution. Generally it is difficult for external people to use the public institution supercomputer casually. However, if it is on a cloud, anyone can use computational resources on thousands of cores instantly when necessary.

Distributed computing has mainly been selected as the method for cloud computing. With the development of grid computing, computation on the cloud by Apache Hadoop has been conducted extensively \cite{Driscoll2013,Karlsson2012,Ekanavake2011,Matsunaga2008} and support tools for constructing Hadoop clusters on the cloud have been established \cite{Hodor2015}. However, while Hadoop/MapReduce can easily construct a distributed task calculation environment, it is versatile and therefore contains an excessive amount of functions. These tools exhibit limited applicability to certain areas such as data mining, because MapReduce provides poor performance on problems with an iterative structure present in the linear algebra that underlies a substantial amount of data analysis \cite{Qiu2010}. To improve the performance and enable flexible design according to scientific applications, an original task distribution system has been constructed based on the message passing interface (MPI) in several cases \cite{Lu2010}. Hassan et al., for example, implemented well-known MPI-based benchmarks (NAS parallel benchmarks) in Azure \cite{Hassan2016}.

Fortunately, AWS and Azure provide instances and networks with awareness of parallel high-performance computing (HPC). For example, in Azure, which was used in this research, an instance of a remote direct memory access (RDMA) network (InfiniBand) is also provided. Such an environment is expected to highly effective for parallel computing applications. However, information such as which instance should be used, the amount of scalability obtained, and the price has not been sufficiently clarified in previous studies. 

Therefore, in this study, a large-scale parallel computation of a bioinformatics application was performed on several cloud instances with suggestions for the choice of the public cloud usage environment. We focused on protein--protein interaction predictions, particularly the protein--protein docking problem, as a bioinformatics application. Protein--protein docking, which is a computational method for predicting the structure of a protein complex from known component structures, is a powerful approach that facilitates the discovery of otherwise unattainable protein complex structures. Fast Fourier transform (FFT)-based rigid-body initial protein--protein docking tools are the mainstream of protein--protein docking (as reviewed by \cite{Matsuzaki2016}). Several applications also require a huge number of dockings, such as consensus-based refinement \cite{Chermak2015,Launey2020}, large-scale interactome predictions \cite{Lopes2013,Ohue2014a,Hayashi2018}, the identification of protein binders \cite{Wass2011,Zhang2014}, and multiple docking \cite{Kraca2011}.
We previously developed the supercomputer-powered software MEGADOCK \cite{Ohue2014a,Matsuzaki2013,Ohue2014b}, and we drew on this experience to develop a protein--protein docking tool for efficient HPC computation on the public cloud.
A protein--protein docking environment that can achieve large-scale analysis on the cloud is necessary in the current global situation, in which large-scale computing environments are readily available on the cloud.

In this study, we demonstrated the implementation and performance of high-performance cloud protein--protein docking. We evaluated the parallelization efficiency (strong scaling) of MEGADOCK implemented on Microsoft Azure, and verified its usage efficiency for GPU instances.

\section{Materials and Methods}
\subsection{Configuration of Azure cloud computing environment}
A unit of computing environment on Azure is called an instance or virtual machine (VM). The machine architecture on Azure is composed of multiple VMs and storage, as illustrated in Figure~1. Each VM and storage is first deployed from AzureCLI and then registered as a resource group in Azure. Thereafter, the computation task is executed on multiple VMs by means of MPI communication. The programs for the bulk VM deployment and bulk undeployment were developed in this study.

\begin{figure}[tb]
  \begin{center}
    \includegraphics[width=\textwidth]{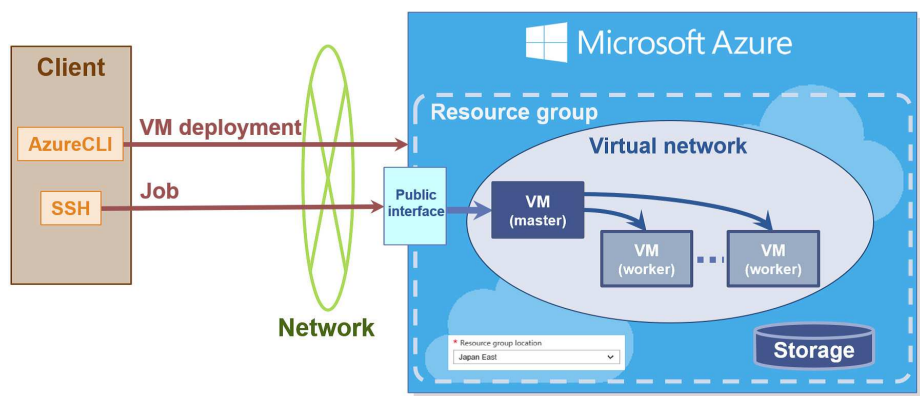}
    \caption{Configuration of Azure cloud computing environment}
    \label{Fig1}
  \end{center}
\end{figure}

\subsection{MEGADOCK: protein--protein docking tool}
MEGADOCK \cite{Ohue2014b} is our software for protein--protein interaction prediction. The 3D structures (PDB data) of two proteins for predicting interaction are input, and presence or absence of the interaction is output in the form of a score. The main part of the calculation is grid-based docking of the protein, which is implemented using FFT \cite{Katchalski1992}. The FFT calculation depends on the protein size, but is approximately 80\%{} of the total occupancy. The computational scale is $O(N^3 \log N)$ if the size of one side of the grid is $N$, usually representing a protein in a grid of 1.2 \AA{} pitches.

MEGADOCK is a multi-threaded implementation that uses OpenMP and runs on a multi-core CPU. Furthermore, a GPU-implemented version is available, which runs on the multiple GPUs using the CUDA library \cite{Shimoda2015}. A multi-node parallel implementation version was also created by hybrid parallelization combined with MPI parallelization \cite{Ohue2014a}. In this work, we constructed parallel implementations for both the CPU VMs and GPU VMs. The details of the parallelization are presented in the following subsection.

\subsection{Handling multiple VMs}
In the multi-node implementation of MEGADOCK, a master--worker-type task dispatching is performed using MPI. Specifically, one process becomes the master process, and tasks are allocated to the worker processes while the remaining tasks and computing resources are monitored. The tasks are independent for each protein pair and can be data parallelized.

In Azure cloud, we adopted the master--worker-type task dispatching in parallel, whereby one process was the master process and the remaining resources were used to execute multiple worker processes, and MPI communication was used to realize the task dispatching for the protein--protein interaction prediction. Unlike the case in a normal cluster-type computing environment, the distance between real machines in a cloud computing environment tends to be large, and MPI implementation is generally not considered as suitable. However, as MEGADOCK does not require heavy communication between tasks (worker processes), it was expected that the large-scale parallelization would not cause serious slowdowns.

Among the Azure instances available for HPC applications, we targeted {\sf A9}, {\sf DS14}, {\sf H16}, and {\sf H16r} as CPU instances with 16 CPU cores, as well as {\sf NC24} and {\sf NC24r} as GPU instances equipped with 24 CPU cores and 4 GPU chips. The details of each instance are displayed in Table~1. For each process to be able to use one GPU, a task dispatching was performed to run four processes per instance (on a VM). That is, the number of CPU cores allocated to each task was 1/4 of the number of cores in each VM: 4 cores for CPU instances and 6 cores for GPU instances.

\begin{table}[tb]
\begin{center}
\caption{Details of Azure instances used in study}
\label{tab1}
\begin{tabular}{c|ccccc} \hline
Instance  & CPU & \#{} cores & Total DP peak (CPU) & \\\hline
{\sf DS14} & Xeon E5-2660 @2.20 GHz $\times$ 2 & 16 & 281.6 GFlops & \\
{\sf A9} & Xeon E5-2670 @2.60 GHz $\times$ 2 & 16 & 332.8 GFlops & \\
{\sf H16} & Xeon E5-2667v3 @3.20 GHz $\times$ 2 & 16 & 691.2 GFlops& \\
{\sf H16r} & Xeon E5-2667v3 @3.20 GHz $\times$ 2 & 16 & 691.2 GFlops& \\\hline
{\sf NC24} & Xeon E5-2690v3 @2.60 GHz $\times$ 2 & 24 & 883.2 GFlops &\\
{\sf NC24r} & Xeon E5-2690v3 @2.60 GHz $\times$ 2 & 24 & 883.2 GFlops &\\\hline\hline
Instance  & GPU & RAM & Network & Price (at March 2017) \\\hline
{\sf DS14}  & N/A & 112 GB & - & 1.39 USD/h \\
{\sf A9} & N/A & 112 GB & RDMA supported & 1.93 USD/h \\
{\sf H16} & N/A & 112 GB & - & 1.75 USD/h \\
{\sf H16r} & N/A & 112 GB & RDMA supported & 1.92 USD/h \\\hline
{\sf NC24} & Tesla K80 $\times$ 4 chips & 1,440 GB & - & 4.32 USD/h \\
{\sf NC24r} & Tesla K80 $\times$ 4 chips & 1,440 GB & RDMA supported & 4.75 USD/h \\\hline
\end{tabular}
\end{center}
\end{table}

\subsection{Experimental settings}
The dataset was the total of 59 protein heterodimeric complexes in the ZLAB protein--protein docking benchmark (version 1.0) \cite{Mintseris2003}. The 59 heterodimers were divided, and all-to-all (cross) docking calculations were performed on the 59 receptor proteins and 59 ligand proteins.

\section{Results and Discussion}
\subsection{MEGADOCK on multiple CPU instances}
The results of the parallel execution of MEGADOCK on 50 and 100 instances using the CPU instances {\sf DS14}, {\sf A9}, {\sf H16}, and {\sf H16r} are presented in Table~2. 
The calculation time values were the median values measured three times. In this case, strong scaling was the value calculated as strong scaling $= (T_{50}/T_{100})/(100/50)$ when the computation times of 50 and 100 instances were $T_{50}$ and $T_{100}$, respectively.

\begin{table}[tb]
\begin{center}
\caption{Results of MEGADOCK on Azure CPU instances (values in parentheses are the ratio of the calculation speed to {\sf H16}.)}
\label{tab2}
\begin{tabular}{c|ll|r} 
\hline
Instance & \multicolumn{1}{c}{50 instances} & \multicolumn{1}{c|}{100 instances} & \multicolumn{1}{c}{Strong scaling}  \\\hline
{\sf DS14} & 3,283 s (0.47)& 1,696 s (0.48)& 0.968  \\
{\sf A9} & 2,369 s (0.64)& 1,352 s (0.61)& 0.876  \\
{\sf H16} & 1,527 s (1)& \ \ \ 820 s (1)& 0.931  \\
{\sf H16r} & 1,640 s (0.93)& \ \ \ 953 s (0.86)& 0.861  \\\hline
\end{tabular}
\end{center}
\end{table}

The experimental results demonstrated that the computation using the {\sf H16} instance was the fastest, followed by {\sf H16r}, {\sf A9}, and {\sf DS14}. This ordering is naturally corresponding to the order of CPU performance (total DP peak) presented in Table~1.
	

When 100 {\sf H16} instances (1,600 CPU cores) were used, the calculation was completed in 820 s. This was the speed at which protein--protein docking calculations could be performed at 255 pairs per minute.

The calculation for the {\sf H16r} instance was slightly slower than that for {\sf H16}.
The {\sf H16r} is an instance that can use the RDMA network interface and exhibits higher communication performance than the {\sf H16}, but MEGADOCK achieves higher performance even without RDMA network.
An RDMA network may not be necessary for many bioinformatics applications in which data parallelization is possible.
Moreover, as an instance with an RDMA network is more expensive than an instance without it, it is more reasonable not to use an RDMA network from a cost perspective.

The strong scaling was greater than 0.85 in the range of this measurement in all instances.

\subsection{MEGADOCK on multiple GPU instances}
\begin{figure}[tb]
  \begin{center}
    \includegraphics[width=\textwidth]{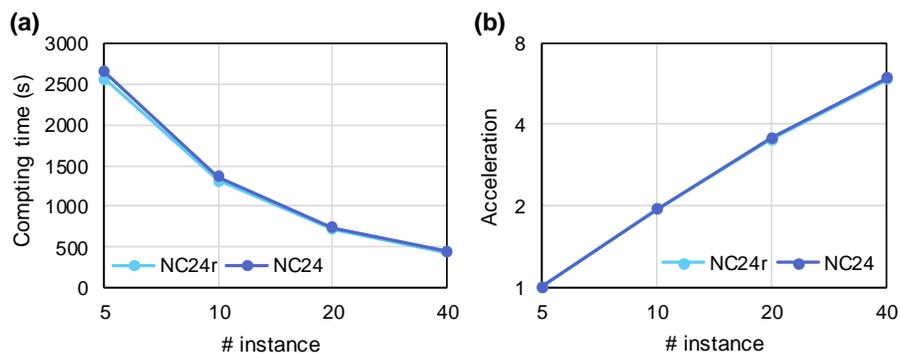}
    \caption{Results of calculation time measurements on GPU instances: (a) computation time for each number of instances and (b) computation speed ratio with respect to five instances}
    \label{Fig2}
  \end{center}
\end{figure}

Using the GPU instances {\sf NC24} and {\sf NC24r}, we measured the computation times with using 5, 10, 20, and 40 instances.
Figure~2 presents the measured calculation times and speed improvement rates.
Owing to the limit of Microsoft Azure on the number of maximum concurrent GPUs (quota limit), the maximum number of allocated instances was 40.
In the comparison between the {\sf NC24} and {\sf NC24r}, the {\sf NC24r} with an RDMA network slightly outperformed the {\sf NC24} in terms of speed, but the difference was very small.
As with the CPU instance, the GPU instance would not require an RDMA network for this application.

{\sf NC24} is discussed below.
When using 40 instances of {\sf NC24} (960 CPU cores and 160 GPUs), the calculation was completed within 448 s.
This was faster than the result for the CPU instance indicated in Table 2 ({\sf H16}: 1,600 CPU cores), and enabled 466 pairs of protein--protein docking to be performed per minute.
For strong scaling, the parallelization efficiency of 20 instances was 0.89 for 5 instances, which was similar to that of the CPU instances.
However, when 40 instances were used, the speed improvement was only 5.91-fold faster than that of 5 instances, with a strong scaling value of 0.74.

\subsection{Which instance should be used from a cost perspective}
\subsubsection*{CPU instance}
According to the comparison of CPU instances, the computation speed of the {\sf H16} instance was the most favorable.
Comparing the {\sf H16} with the less expensive {\sf DS14}, the speed improvement ratio was 1,696 s/820 s = 2.07. The price ratio between {\sf H16} (1.75 USD/h) and {\sf DS14} (1.39 USD/h) was 1.75 USD/ 1.39 USD = 1.26. 
As a result, it is more reasonable to use the {\sf H16} than the {\sf DS14}, as the value of the speed improvement ratio is larger than the price ratio.

Both {\sf A9} and {\sf H16r} are slightly more expensive because they have an RDMA network, but MEGADOCK does not need to use these instances because no increase obtained in the computation speed when using an RDMA network.
When using applications that require a powerful network, we recommend the {\sf H16r}, which is approximately the same price as the {\sf A9}, but provides higher CPU performance.

\subsubsection*{GPU instance}
A significant increase in the speed was achieved when using the GPU instance.
However, unlike the {\sf H16} and {\sf DS14}, the {\sf NC24} has 24 CPU cores, making a direct comparison difficult.
In the following, we consider the maximum measurements at {\sf H16} (100 instances, 1,600 cores, and 820 s) and {\sf NC24} (40 instances, 960 cores and 160 GPUs, and 448 s) in terms of the cost.
Table 3 provides a summary of these results.
In Table~3, the total fee was calculated by ignoring the time required for factors such as VM deployment and assuming that the product of \{calculation time $\times$ number of instances\} used was the total cloud usage time.

Consequently, the same calculation could be performed for 21.5 USD for {\sf NC24}, compared to 39.9 USD for {\sf H16}.
The {\sf NC24} has a shorter execution time and is almost twice as advantageous in terms of usage fees.
For GPU-enabled applications, the use of GPU instances offers the potential to yield computational results rapidly and inexpensively, and active consideration thereof is recommended.

\begin{table}[tb]
\begin{center}
\caption{Summary of results for {\sf H16} and {\sf NC24} instances}
\label{tab3}
\begin{tabular}{c|ccccc|c} \hline
Instance & \#{} inst. & CPU cores & GPUs & Time & Price (1 inst.) & Total fee$^*$ \\\hline
{\sf H16} & 100 & 1,600 & N/A & 820 s & 1.75 USD/h & 39.9 USD \\
{\sf NC24} & 40 & 960 & 160 & 448 s & 4.32 USD/h & 21.5 USD \\\hline
\end{tabular}\\
$^*$ The total fee was obtained by Price $\times$ Time (h) $\times$ \#{} inst. \quad \quad \quad \quad
\end{center}
\end{table}

\section{Conclusions}
We constructed a computing environment for large-scale protein--protein docking calculations with the MEGADOCK software on the public cloud of Microsoft Azure, and performed large-scale parallel calculations on approximately 1,000 GPUs. We found that MEGADOCK provided the fastest GPU computation on the {\sf NC24} instance and the cloud computing cost was lower than that of using CPU instances. 

Large-scale data analysis with MEGADOCK requires high CPU and GPU performance, but does not require high communication performance.
For bioinformatics applications similar in properties to MEGADOCK, it would be most cost-effective to use the {\sf NC24} instance or the similar instance without high-bandwidth network, like RDMA, as in this study.

The use of the public cloud environment is advantageous owing to the portability and reproducibility of computing applications, and it allows for the rapid construction of large-scale applications such as the one investigated in this study. The cloud environment is particularly useful in applications such as pipeline software, in which various tools are intricately interrelated. We have already developed a system to enable MEGADOCK computation by constructing a container virtualization environment on the cloud \cite{Aoyama2019}. In addition to the protein--protein docking calculations demonstrated in this study, various other bioinformatics applications operating on the public cloud will certainly contribute to accelerating the research in this field.

\section*{Acknowledgements}
The authors thank Mr. Hiroyuki Sato at IMSBIO, Co., Ltd. for his technical support in the development on Microsoft Azure. 
This work was partially supported by the Japan Society for the Promotion of Science (JSPS) KAKENHI (18K18149, 20H04280), Core Research for Evolutional Science and Technology (CREST) ``Extreme Big Data'' (Grant No. JPMJCR1303) from the Japan Science and Technology Agency (JST), the Platform Project for Supporting Drug Discovery and Life Science Research (Basis for Supporting Innovative Drug Discovery and Life Science Research (BINDS)) (Grant No. JP18am0101112) from the Japan Agency for Medical Research and Development (AMED), Microsoft Business Investment Funding from Microsoft Corporation, and Leave a Nest Grant from Leave a Nest Co., Ltd. This work was partially conducted as research activities of AIST-Tokyo Tech Real
World Big-Data Computation Open Innovation Laboratory (RWBC-OIL). The authors thank Editage (www.editage.com) for English language editing.

%
%

\end{document}